\documentclass[11pt]{article}

\usepackage[numbers]{natbib}
\usepackage[utf8]{inputenc}
\usepackage[english]{babel}
\usepackage{graphicx}         
\usepackage[pdftex, colorlinks=true, linkcolor=blue, citecolor=blue, urlcolor=black]{hyperref}
\usepackage{braket}
\usepackage{amsmath}
\usepackage{amssymb}
\usepackage{amsthm}
\usepackage{mathtools}
\usepackage{bbm}		
\usepackage{mathrsfs}
\usepackage{array}		
\usepackage{color}
\usepackage[usenames,dvipsnames]{xcolor}
\usepackage{caption}
\usepackage{enumitem}
\usepackage{subeqnarray}	
\usepackage{wrapfig}

\usepackage{tikz}
\usetikzlibrary{arrows, decorations.markings}

\allowdisplaybreaks

\theoremstyle{plain}

\theoremstyle{remark}

\theoremstyle{plain}

\theoremstyle{plain}

\theoremstyle{plain}\newtheorem{theorem}{Theorem}

\newcommand{\R}{\mathbb{R}}

\newcommand{\N}{\mathbb{N}}

\newcommand{\Fock}{\mathcal{F}}
\newcommand{\Number}{\mathcal{N}}
\newcommand{\vac}{|\Omega\rangle}
\newcommand{\id}{\mathbbm{1}}

\renewcommand{\i}{\mathrm{i}}
\newcommand{\e}{\mathrm{e}}

\newcommand{\sym}{\mathrm{sym}}

\newcommand{\Tr}{\mathrm{Tr}}

\renewcommand{\hat}[1]{\widehat{#1}}
\renewcommand{\tilde}[1]{\widetilde{#1}}

\newcommand{\lr}[1]{\left\langle #1 \right\rangle}
\newcommand{\norm}[1]{\lVert#1\rVert}

\newcommand{\scp}[2]{\langle #1, #2 \rangle}

\renewcommand{\d}{\mathop{}\!\mathrm{d}}
\newcommand{\dx}{\d x}
\newcommand{\dy}{\d y}
\newcommand{\dz}{\d z}

\newcommand{\ad}{a^*}

\newcommand\mydots{,\makebox[1em][c]{.\hfil.\hfil.},}
\newcommand\mycdots{\makebox[1em][c]{$\cdot$\hfil$\cdot$\hfil$\cdot$}}

\makeatletter
\DeclareFontFamily{OMX}{MnSymbolE}{}
\DeclareSymbolFont{MnLargeSymbols}{OMX}{MnSymbolE}{m}{n}
\SetSymbolFont{MnLargeSymbols}{bold}{OMX}{MnSymbolE}{b}{n}
\DeclareFontShape{OMX}{MnSymbolE}{m}{n}{
    <-6>  MnSymbolE5
   <6-7>  MnSymbolE6
   <7-8>  MnSymbolE7
   <8-9>  MnSymbolE8
   <9-10> MnSymbolE9
  <10-12> MnSymbolE10
  <12->   MnSymbolE12
}{}
\DeclareFontShape{OMX}{MnSymbolE}{b}{n}{
    <-6>  MnSymbolE-Bold5
   <6-7>  MnSymbolE-Bold6
   <7-8>  MnSymbolE-Bold7
   <8-9>  MnSymbolE-Bold8
   <9-10> MnSymbolE-Bold9
  <10-12> MnSymbolE-Bold10
  <12->   MnSymbolE-Bold12
}{}
\let\llangle\@undefined
\let\rrangle\@undefined
\DeclareMathDelimiter{\llangle}{\mathopen}
                     {MnLargeSymbols}{'164}{MnLargeSymbols}{'164}
\DeclareMathDelimiter{\rrangle}{\mathclose}
                     {MnLargeSymbols}{'171}{MnLargeSymbols}{'171}
\makeatother

\newcommand\smallO[1]{
        \mathchoice
            {
                \ensuremath{\mathop{}\mathopen{}{\scriptstyle\mathcal{O}}\mathopen{}\left(#1\right)}
            }
            {
                \ensuremath{\mathop{}\mathopen{}{\scriptstyle\mathcal{O}}\mathopen{}\left(#1\right)}
            }
            {
                \ensuremath{\mathop{}\mathopen{}{\scriptscriptstyle\mathcal{O}}\mathopen{}\left(#1\right)}
            }
            { 
                \ensuremath{\mathop{}\mathopen{}{o}\mathopen{}\left(#1\right)}
            }
    }

\newcommand{\HN}{H_N}
\newcommand{\Vext}{V^\mathrm{trap}}

\newcommand{\PsiN}{\Psi_N}

\newcommand{\EN}{\mathscr{E}_N}

\newcommand{\goN}{\gamma_N^{(1)}}

\newcommand{\eH}{e_\mathrm{H}}

\newcommand{\cEH}{\mathcal{E}_\mathrm{H}}

\newcommand{\UNp}{U_{N,\varphi}}

\newcommand{\Fp}{{\Fock_{\perp\varphi}}}
\newcommand{\FNp}{{\Fock_{\perp\varphi}^{\leq N}}}

\newcommand{\FockH}{\mathbb{H}}
\newcommand{\FockHz}{\FockH_0}
\newcommand{\FockHo}{\FockH_1}
\newcommand{\FockHt}{\FockH_2}
\newcommand{\FockHj}{\FockH_j}

\newcommand{\FockR}{\mathbb{R}}

\newcommand{\Chi}{{\boldsymbol{\chi}}}
\newcommand{\Chiz}{\Chi_0}

\renewcommand{\P}{\mathbb{P}}
\newcommand{\Q}{\mathbb{Q}}

\newcommand{\Pz}{\P_0}
\newcommand{\Qz}{\Q_0}

\newcommand{\Ez}{E_0}

\newcommand{\Np}{\Number_{\perp\varphi}}

\newcommand{\BogUz}{\mathbb{U}_0}

\newcommand{\pt}{{\varphi(t)}}

\newcommand{\mpt}{\mu^{\pt}}

\newcommand{\cBN}{\mathcal{B}_N}

\title{Asymptotic analysis of the weakly interacting  Bose gas:\\ A collection of recent results and applications}
\author{Lea Boßmann\thanks{Mathematisches Institut, Ludwig-Maximilians-Universität München, Theresienstr.\ 39, 80333 München, Germany. Email: \texttt{bossmann@math.lmu.de}},
Nikolai Leopold\thanks{Department of Mathematics and Computer Science, University of Basel, Spiegelgasse 1, 4051 Basel, Switzerland. Email: \texttt{nikolai.leopold@unibas.ch}},
David Mitrouskas\thanks{Institute of Science and Technology Austria (ISTA), Am Campus 1, 3400 Klosterneuburg, Austria. Email: \texttt{mitrouskas@ist.ac.at}}, and
Sören Petrat\thanks{School of Science, Constructor University, Campus Ring 1, 28759 Bremen, Germany. Email: \texttt{spetrat@constructor.university}}}
\date{\today}

\begin{document}
\maketitle

\begin{center}
\emph{To Detlef Dürr, \\ a great teacher and wonderful person.}
\end{center}

\vspace{1mm}

\frenchspacing

\begin{abstract}
We consider a gas of $N$ bosons with interactions in the mean-field scaling regime.
We review a recent proof of the asymptotic expansion of its spectrum and eigenstates and two applications of this result, namely the derivation of an Edgeworth expansion for fluctuations of one-body operators and the computation of the binding energy of an inhomogeneous Bose gas to any order.
Finally, we collect related results for the dynamics of the weakly interacting Bose gas and for the regularized Nelson model.
\end{abstract}

\section{Introduction}

Bose gases have been studied from many different perspectives since the discovery of Bose--Einstein condensation (BEC), which, after the theoretical prediction in 1924 by Bose \cite{Bose_original} and Einstein \cite{Einstein_original_1,Einstein_original_2}, was first experimentally realized in 1995 by the groups of Cornell/Wieman \cite{anderson1995} and Ketterle \cite{davis1995}. In a typical experiment, the bosons are initially caught in an external trap, where they are cooled down to a superposition of low-energy eigenstates; subsequently, they are released and their behavior is observed. If the number of particles in the gas is large, neither an  analytical nor a numerical analysis of the system is feasible, which makes the use of appropriate approximations indispensable.

The resulting evolution equations are sometimes broadly called \emph{effective equations}. The study of their emergence from a microscopic theory of interacting particles is a typical question in mathematical and statistical physics. In a different context, namely that of conductivity, and also of Brownian motion, this field is where Detlef started his career as a mathematical physicist. We therefore like to think that he would have enjoyed the kind of results we are presenting here, and we dedicate this article to him.

Over the last two decades in particular, there have been many contributions in the mathematical physics community devoted to a rigorous derivation of suitable effective equations for different models of BEC. In this review, we restrict ourselves to the weakly interacting Bose gas, also known as the mean-field or Hartree regime, which describes trapped bosons with weak and long-range interactions. The corresponding Hamiltonian for the $N$-body system is given by
\begin{equation}\label{HNtrap}
\HN=\sum\limits_{j=1}^N\left(-\Delta_j+\Vext(x_j)\right)+\frac{1}{N-1}\sum\limits_{1\leq i<j\leq N}v(x_i-x_j)\,,
\end{equation}
acting on the Hilbert space $L^2_\sym((\R^d)^N)$ of square integrable, permutation symmetric functions on $(\R^d)^N$.  
We assume the two-body interaction potential $v:\R^d\to\R$ to be bounded, symmetric and---for our spectral results---of positive type, i.e., to have a non-negative Fourier transform.
The confining potential $\Vext:\R^d\to\R$ is assumed to be measurable, locally bounded, non-negative, and such that $\Vext(x)$ tends to infinity as $|x|\to  \infty$. Instead of using an external potential in $\R^d$, one often restricts the particles to the $d$-dimensional unit torus $\mathbb{T}^d$, which usually simplifies the analysis since the resulting system is homogeneous.

The spectral and dynamical properties of the model \eqref{HNtrap} have been subject to extensive research; for more recent results, see, e.g., \cite{seiringer2011, grech2013, lewin2014, lewin2015_2, mitrouskas_PhD} and \cite{grillakis2010,grillakis2011,lewin2015,nam2015,mitrouskas2016}, respectively. Let us also refer to \cite{LSSY} for a more general review of BEC.

In this article, we start in Section~\ref{sec_spectrum} by reviewing results related to the spectrum and eigenfunctions based on \cite{spectrum}. In Section~\ref{sec_applications}, we review the Edgeworth expansion from \cite{CLT} and the binding energy expansion from \cite{binding_energy_reference}. Finally, in Section~\ref{sec:dynamics}, we review the dynamical results from \cite{QF} and \cite{FLMP2021}.

In the following Sections~\ref{sec_spectrum} and \ref{sec_applications}, we consider the ground state $\PsiN$ of $\HN$ and the ground state energy $\EN$, i.e.,
\begin{equation}
\EN=\inf\mathrm{spec}(\HN)\,,\qquad \HN\PsiN=\EN\PsiN\,.
\end{equation}
Under appropriate conditions on $v$ and $\Vext$, it is well known that $\PsiN$ is unique and exhibits complete asymptotic BEC in the minimizer $\varphi \in L^2(\mathbb R^d)$ of the Hartree energy functional, which is given by
\begin{equation}\label{def:Hartree:functional}
\cEH[\phi]:=\int\limits_{\R^d}\left(|\nabla \phi(x)|^2+\Vext(x)|\phi(x)|^2\right)\dx
+\tfrac12\int\limits_{\R^{2d}}v(x-y)|\phi(x)|^2|\phi(y)|^2\dx\dy\,.
\end{equation}
We denote its minimum under the constraint $\norm{\phi}=1$ by $\eH:=\cEH[\varphi]$.
Complete asymptotic BEC in the state $\varphi$ means that $\PsiN$ is determined by $\varphi$ in the sense of reduced densities, i.e., 
\begin{equation}\label{eqn:BEC}
\lim\limits_{N\to\infty} \Tr\,\left|\goN-|\varphi\rangle\langle\varphi|\right| =0\,,
\end{equation}
where $\goN:=\Tr_{2\mydots N}|\PsiN\rangle\langle\PsiN|$ denotes the one-particle reduced density matrix of $\PsiN$. Heuristically, this implies that $N-\smallO{N}$ particles occupy the condensate state $\varphi$. Consequently, the leading order of $\EN$ is given by the condensate energy $N\eH$.

\section{Asymptotic expansion of the ground state}\label{sec_spectrum}
\subsection{Main result}
The first result we review in these notes is an expansion of the $N$-body ground state $\PsiN$ and of the ground state energy $\EN$ in powers of $N^{-1/2}$, which is proven in \cite{spectrum}.

\begin{theorem}\label{thm:eigenstates:spectrum}
Let $a\in\N_0$ and let $N$ be sufficiently large. Then there exists a constant $C(a)$ such that
\begin{equation}\label{eqn:thm:eigenstates}
\Big\|\PsiN-\sum\limits_{\ell=0}^a N^{-\frac{\ell}{2}}\psi_{N,\ell}\Big\|_{L^2((\R^d)^N)}\leq C(a)N^{-\frac{a+1}{2}}
\end{equation}
and
\begin{equation}\label{eqn:thm:spectrum}
\left|\EN-  N\eH -  \sum\limits_{\ell=0}^a N^{-\ell} E_\ell \right|\; \leq\; C(a) N^{-(a+1)}\,.
\end{equation}
The coefficients  $\psi_{N,\ell}\in L^2_\sym((\R^d)^N)$ and $E_\ell \in\R$ are computed in \cite{spectrum} in full generality. As an example, $\psi_{N,1}$ and $E_1$ are given in \eqref{Chi_1} and \eqref{E_1}.
\end{theorem}
To leading order ($a=0$), this was proven in \cite{seiringer2011,grech2013,lewin2015_2,mitrouskas_PhD}. The higher orders $(a>0)$ were  rigorously derived in \cite{spectrum}, and related results were obtained in \cite{pizzo2015,pizzo2015_2,pizzo2015_3}.

The coefficients $\eH$ and $E_\ell$ are independent of $N$. The $N$-body wave functions $\psi_{N,\ell}$ naturally depend on $N$; however, this $N$-dependence is  trivial, which is explained below. As a result, the computational effort to obtain physical quantities such as expectation values with respect to the $N$-body state, does not scale with $N$. 

The constants $C(a)$ grow rapidly in $a$, which means that \eqref{eqn:thm:eigenstates} and \eqref{eqn:thm:spectrum} are \emph{asymptotic} expansions (and not converging series): given any order $a$ of the approximation, one can choose $N$ sufficiently large that the estimates are meaningful.

Theorem \ref{thm:eigenstates:spectrum} extends to the low-energy excitation spectrum of $\HN$ and to a certain class of unbounded interaction potentials $v$, including the repulsive three-dimensional Coulomb potential (see \cite{spectrum} for the full statement). Moreover, it implies an asymptotic expansion of the corresponding one-body reduced density matrices \cite{proceedings}.

\subsection{Idea of proof}
The contributions to the ground state energy beyond the leading order are caused by particles  which are excited from the condensate due to the interactions.
To describe these excitations, one decomposes $\PsiN$ as
\begin{equation}\label{eqn:decomposition:PsiN}
\PsiN=\sum\limits_{k=0}^N{\varphi}^{\otimes (N-k)}\otimes_s\chi^{(k)}\,,
\qquad \chi^{(k)}\in \bigotimes\limits_\sym^k \{\varphi\}^\perp\,, \qquad
\Chi:=\big(\chi^{(k)}\big)_{k=0}^N\in\FNp\subset\Fp
\end{equation}
with $\otimes_s$ the symmetric tensor product and where $\{\varphi\}^\perp$ denotes the orthogonal complement of $\varphi$ in $L^2(\R^d)$ \cite{lewin2015_2}.  
The excitations
form a vector in the (truncated) excitation Fock space over $\{\varphi\}^\perp$, which is denoted by $\Fp$ (resp. $\FNp$).
The creation/annihilation operators $\ad$/$a$ and the number operator $\Np$ on this Fock space are defined in the usual way.
The relation between $\PsiN$ and the corresponding excitation vector $\Chi$ is given by the unitary map
\begin{eqnarray}\label{map:U}
\UNp:L^2((\R^d)^N) \to  \FNp\;, \quad
 \PsiN  \mapsto  \UNp \PsiN= \Chi\,.
\end{eqnarray}
Conjugating $\HN$ with $\UNp$ and subtracting the condensate energy $N\eH$ yields the operator 
\begin{equation}\label{FockH}
\FockH\;:=\;\UNp\left(\HN-N\eH\right)\UNp^*
\end{equation}
on $\FNp$, whose ground state is denoted by $\Chi$. Hence, the ground state energy $E$ of $\FockH$,
\begin{equation}
E=\lr{\Chi,\FockH\Chi}_{\FNp}=\EN-N\eH\,,
\end{equation} 
gives us precisely the corrections to the condensate energy $N\eH$ in \eqref{eqn:thm:spectrum}.
After extending $\FockH$ trivially to the full excitation Fock space $\Fp$, computing \eqref{FockH} as in \cite[Proposition 4.2]{lewin2015_2} yields an expansion of $\FockH$ in powers of $N^{-1/2}$,
\begin{equation}\label{intro:taylor}
\FockH \;=\;\FockHz+\sum\limits_{j= 1}^aN^{-\frac{j}{2}}\FockHj+N^{-\frac{a+1}{2}}\FockR_a
\end{equation}
for any $a\in\N_0$.
The coefficients $\FockHj$ and the remainders $\FockR_a$ in this expansion are unbounded operators on $\Fp$ which depend on $v$, $\Vext$ and $\varphi$. The operators $\FockHj$ are independent of $N$.

The leading order term $\FockHz$ in \eqref{intro:taylor} is the well-known Bogoliubov Hamiltonian, which is a very useful approximation of $\FockH$ because it is quadratic in the number of creation/annihilation operators. Under the given assumptions on $v$, it can therefore be diagonalized by a Bogoliubov transformation $\BogUz$, in the sense that $\BogUz \FockHz \BogUz^* = E_0 + \int \dx \,a_x^* D(x,y) a_y$ for some positive one-body operator $D$.
The unique ground state of $\FockHz$ is thus given by
\begin{equation}\label{eqn:Chiz}
\Chiz=\BogUz^*\vac\,,
\end{equation}
where $\vac$ is the vacuum state, and its ground state energy is $\Ez$. It is well known \cite{seiringer2011,grech2013,lewin2015_2,mitrouskas_PhD} that 
\begin{equation}\label{eqn:known:results:FockHN:E}
\lim\limits_{N\to\infty}E=\lim\limits_{N\to\infty}(N\eH-\EN)=\Ez\,,\qquad
\lim\limits_{N\to\infty}\norm{\Chi-\Chiz}_{\Fp}=0\,,
\end{equation}
where we trivially extended $\Chi$ to a vector in $\Fp$.
Consequently, $\Ez$ gives the leading order ($a=0)$ correction to $\mathcal{E}_N-N\eH$ in \eqref{eqn:thm:spectrum}; analogously, the leading order contribution in \eqref{eqn:thm:eigenstates} is given by $\psi_{N,0}=\UNp^*\,\Chiz|_{\FNp}$.
\medskip

Assuming that $\Chi$ and $E$ have expansions in $N^{-1/2}$, an expansion of the eigenvalue equation yields
\begin{equation}\label{eq_EV_order_1}
(\FockHz - E_0) \Chi_1 + (\FockH_1 - E_{1/2}) \Chi_0 = 0,
\end{equation}
where $E_{1/2}$ is the coefficient of $N^{-1/2}$ in the expansion of $E$. Projecting this equation on $\Chi_0$ with the projector $\Pz:=|\Chiz\rangle\langle\Chiz|$ and then using $\FockHz \Chi_0 = E_0 \Chi_0$, we find
\begin{equation}\label{E_1_2_vanishes}
E_{1/2} = \langle \Chi_0, \FockH_1 \Chi_0 \rangle = 0,
\end{equation}
where the last equality follows since $\FockH_1$ is cubic in the number of creation and annihilation operators and $\BogUz$ is a Bogoliubov transformation, i.e., it maps linear combinations of $\ad / a$ into linear combinations of $\ad / a$. (Alternatively, one can argue that $\Chi_0$ is quasi-free, and thus the left-hand side of \eqref{E_1_2_vanishes} vanishes due to Wick's rule.) Therefore, no $N^{-1/2}$ order appears in the energy expansion \eqref{eqn:thm:spectrum}; in fact, similar arguments can be used to show that every half-integer power of $N^{-1}$ vanishes. Projecting Equation~\eqref{eq_EV_order_1} on the orthogonal complement using $\Qz = 1-\Pz$, we find
\begin{align}\begin{split}\label{Chi_1}
\Chi_1 &=\frac{\Qz}{\Ez-\FockHz}\FockH_1\Chiz = \BogUz^* \left(\BogUz\frac{\Qz}{\Ez-\FockHz}\BogUz^*\right) \BogUz\FockH_1\BogUz^* |\Omega\rangle \\
&= \BogUz^*\left(\int_{\mathbb{R}^d} \dx\,\Theta_1(x)\ad_x|\Omega\rangle + \int_{\mathbb{R}^{3d}} \dx^{(3)}\Theta_3(x^{(3)})\ad_{x_1}\ad_{x_2}\ad_{x_3}|\Omega\rangle\right)\,,
\end{split}\end{align}
where we abbreviate $x^{(3)} = (x_1,x_2,x_3)$. Note that the last equality follows again from the facts that  $\FockH_1$ is cubic in $\ad / a$ and that $\BogUz$ is a Bogoliubov transformation, as well as using that $\BogUz\frac{\Qz}{\Ez-\FockHz}\BogUz^*$ is particle-number conserving; the functions $\Theta_1\in L^2(\R^d)$ and $\Theta_3\in L^2((\R^d)^3)$ can then be explicitly computed. Finally, the coefficients $\psi_{N,\ell}$ in the expansion \eqref{eqn:thm:eigenstates} of the $N$-body ground state $\PsiN$  (Theorem~\ref{thm:eigenstates:spectrum}) are constructed from \eqref{Chi_1} by \eqref{eqn:decomposition:PsiN}. The functions $\psi_{N,\ell}$ depend on $N$ by construction. However, this $N$-dependence is trivial, since it comes only from the splitting into condensate $\varphi$ and excitations $\Chi$. The coefficients $\Chi_\ell$ in the expansion of $\Chi$ are completely independent of $N$.

To prove Theorem~\ref{thm:eigenstates:spectrum}, we follow a different route than using the eigenvalue equation: we expand $\P:=|\Chi\rangle\langle\Chi|$ around $\Pz$ in a (Rayleigh-Schrödinger) perturbation series. By \eqref{eqn:known:results:FockHN:E}, the projectors $\P$ and $\Pz$ can be expressed as
\begin{equation}
\Pz=\frac{1}{2\pi\i}\oint_\gamma\frac{1}{z-\FockHz}\dz\,,\qquad
\P=\frac{1}{2\pi\i}\oint_\gamma\frac{1}{z-\FockH}\dz\,,
\end{equation}
for any $\mathcal{O}(1)$-contour $\gamma$ whose interior contains both $E$ and $\Ez$ but no other point from the spectra of $\FockH$ and $\FockHz$; this is possible by \eqref{eqn:known:results:FockHN:E} and since $\FockH$ and $\FockHz$ have a spectral gap of $\mathcal{O}(1)$. Now, one uses the expansion \eqref{intro:taylor} of $\FockH$ to expand $(z-\FockH)^{-1}$ around $(z-\FockHz)^{-1}$. 
Since $\P$ is a rank-one projector, this immediately implies an expansion of the corresponding vector $\Chi$. After some lengthy computations using \eqref{intro:taylor}, the identity \eqref{eqn:Chiz}, the fact that $\FockH_j$ for $j$ odd (even) is odd (even) in the number of creation and annihilation operators, and that $\BogUz$ is a Bogoliubov transformation diagonalizing $\FockHz$, one obtains the expansion \eqref{Chi_1} and the higher orders by using Cauchy's integral formula.

\medskip
The main work in the proof of Theorem \ref{thm:eigenstates:spectrum} is to estimate the error terms in the expansions above. For example, to control the error for $a=1$, we bound $\FockH_1$, $\FockR_0$ and $\FockR_1$ by powers of $(\Np+1)$, prove a uniform bound on finite moments of the number operator with respect to $\Chi$, and provide suitable estimates for the commutators of powers of $\Np$ with resolvents of $\FockHz$. The expansion of the ground state energy $\mathcal{E}_N$ is then another consequence of the expansion of $\P$. For example, the next order term after the Bogoliubov energy is given by
\begin{eqnarray}\label{E_1}
E_1&=& \lr{\Chiz,\FockHt\Chiz} +\lr{\Chiz,\FockHo\frac{\Qz}{\Ez-\FockHz}\FockHo\Chiz}
\,.
\end{eqnarray}

\section{Applications}\label{sec_applications}
\subsection{Edgeworth expansion}

Let the  Bose gas be in its ground state $\PsiN$ and consider the statistics of experiments described by self-adjoint one-body operators on $L^2((\R^d)^N)$, i.e., operators of the form
\begin{equation}
B_j=\underbrace{\id\otimes\mycdots\otimes\id}_{j-1}\otimes B\otimes \underbrace{\id\otimes\mycdots\otimes\id}_{N-j}\,.
\end{equation}
By the Born rule and since $\PsiN$ is permutation symmetric, the family $\{B_j\}_{j=1}^N$ defines a family of identically distributed random variables: the probability that the random variable $B_j$ takes values in $A\subset\R$ is given by
\begin{equation}
\mathcal{P}_{\PsiN}(B_j\in A)=\lr{\PsiN,\id_A(B_j)\PsiN}\,,
\end{equation}
where $\id_A$ denotes the characteristic function of the set $A$. Since we consider $N$ indistinguishable bosons, we are interested in describing the statistics of experiments described by symmetrized operators $\sum_{j=1}^N B_j$. Centering and rescaling leads us to consider operators
\begin{equation}\label{B_N_RV}
\cBN:=\frac{1}{\sqrt{N}}\sum_{j=1}^N(B_j-\mathbb{E}_{\PsiN}[B])\,,
\end{equation}
where $\mathbb{E}_{\PsiN}[B]=\lr{\PsiN,B_1\PsiN}$.
From Theorem~\ref{thm:eigenstates:spectrum} we know that $\PsiN$ is not a product state, which implies that the random variables $B_j$ are not independent. However, their dependency is weak, and on the level of the excitation Fock space, the correlations are described to leading order by a quasi-free state, i.e., a Bogoliubov transformation acting on the vacuum as in \eqref{eqn:Chiz}. Quasi-free states satisfy a Wick rule in analogy to Gaussian random variables, hence to leading order the statistics of \eqref{B_N_RV} can be expected to be Gaussian. Indeed, it is shown in \cite{CLT} that the fluctuations satisfy a weak Edgeworth expansion:

\begin{theorem}\label{thm:edgeworth}
Let $a\in\N_0$ and $g\in L^1(\R)$ such that its Fourier transform $\hat{g}\in L^1(\R,(1+|k|^{3a+4}))$.
Then, for any self-adjoint bounded operator $B$ on $L^2(\R^d)$, there exists $C_B(a,g)>0$ such that
\begin{equation}\label{eqn:thm:edgeworth}
\left|\mathbb{E}_{\PsiN}[g(\mathcal{B}_N)]
- \sum_{j=0}^aN^{-\frac{j}{2}}
\int_{\mathbb{R}^d} \dx\, g(x){p}_j(x) \, \frac{1}{\sqrt{2\pi\sigma^2}} \e^{-\frac{x^2}{2\sigma^2}} \right|
\leq C_B(a,g) N^{-\frac{a+1}{2}}\,.
\end{equation}
The functions ${p}_j$ are real polynomials of degree $3j$ which are even/odd for $j$ even/odd. In particular,
\begin{subequations}
\begin{eqnarray}
{p}_0 (x)&=&1\,,\\
{p}_1(x) &=& \frac{\alpha}{6\sigma^3}H_3\left(\frac{x}{\sigma}\right)\,,
\end{eqnarray}
\end{subequations}
where $H_3(x)=x^3-3x$ is the third Hermite polynomial. The $N$-independent parameters $\sigma,\alpha\in\R$ are  given in \eqref{sigma} and in \cite{CLT}.
\end{theorem}

The leading order ($a=0$) of the expansion is a central limit theorem, which was proven in \cite{rademacher2019} (see also \cite{benarous2013} for the related dynamical result). 
Analogously to Theorem \ref{thm:eigenstates:spectrum}, the constant $C_B(g,a)$ in Theorem \ref{thm:edgeworth} grows in $a$, hence \eqref{eqn:thm:edgeworth} is an asymptotic expansion. 
It constitues a weak Edgeworth expansion in the sense of \cite{fernando2021}, which, in particular, does not imply an asymptotic expansion of the probability $\mathcal{P}_{\PsiN}(\cBN\in A)$ for $A\subset\R$. Also note that Edgeworth expansions give us a detailed picture of the probability distribution near the expectation value. A more detailed description of the tails are large deviation results, see, e.g., \cite{kirkpatrick2021,rademacher2021_2}.

Theorem \ref{thm:edgeworth} extends to a class of low-energy excited states of $\HN$. In this case, one does not obtain a Gaussian central limit theorem, because these excited states are not quasi-free. However, they are still given by some polynomial of creation operators acting on a quasi-free state, hence the limiting distribution is a Gaussian multiplied with a polynomial. This leads to a generalized Edgeworth-type expansion with different polynomials of higher degree (see \cite{CLT} for the details).
\medskip

To prove Theorem \ref{thm:edgeworth}, we show an expansion of the characteristic function of the random variable $\cBN$. Making use of the expansion $\Chi=\Chiz+\mathcal{O}(N^{-1/2})$ from Theorem \ref{thm:eigenstates:spectrum}, we obtain 
\begin{eqnarray}\label{expansion:char:fctn}
\lr{\PsiN,\e^{\i k\cBN}\PsiN}
&=&\lr{\Chi,\e^{\i k\UNp\cBN\UNp^*}\Chi}
\,=\,\lr{\Omega,\e^{\i k \BogUz(\ad(qB\varphi)+a(qB\varphi))\BogUz^*}\Omega}+\mathcal{O}(N^{-\frac12})\nonumber\\
&=&\e^{-\frac12\sigma^2k^2}+\mathcal{O}(N^{-\frac12})
\end{eqnarray}
with $q:=1-|\varphi\rangle\langle\varphi|$ and where
\begin{equation}\label{sigma}
\sigma:=\norm{\nu}\,,\qquad \nu:=U_0qB\varphi+\overline{V_0qB\varphi}\,.
\end{equation}
for certain bounded operators $U_0,V_0 $ on $\{\varphi\}^\perp\subset L^2(\R^d)$.
Here, we used that
\begin{equation}
\BogUz\left(\ad(qB\varphi)+a(qB\varphi)\right)\BogUz^*=\ad(\nu)+a(\nu)
\end{equation}
since $\BogUz$ from \eqref{eqn:Chiz} is a Bogoliubov transformation. By Fourier transformation, this yields \eqref{eqn:thm:edgeworth} for $a=0$. The higher orders in \eqref{eqn:thm:edgeworth} are computed along the same lines, making use of higher orders in Theorem \ref{thm:eigenstates:spectrum}.
\medskip

Finally, let us compare Theorem \ref{thm:edgeworth} (which concerns the fluctations $\cBN$ of dependent random variables distributed according to $\PsiN$) with the corresponding result for the fluctuations $\cBN^\mathrm{iid}$ of i.i.d.\ random variables distributed according to the product state $\varphi^{\otimes N}$. Standard probability theory (e.g.\ \cite{petrov}) yields for $\cBN^\mathrm{iid}$ an Edgeworth expansion which is of the same structure as \eqref{eqn:thm:edgeworth}, i.e., a Gaussian multiplied with polynomials of degree $3j$ with the same even/odd structure. However, there are important differences: First, the variance of the Gaussian in the i.i.d.\ case is given by $\sigma^2_\mathrm{iid}=\norm{qB\varphi}^2=\lr{\varphi,B^2\varphi}-\lr{\varphi,B\varphi}^2\neq\sigma^2$, which can be seen analogously to \eqref{expansion:char:fctn} noting that $\UNp\varphi^{\otimes N}=|\Omega\rangle$. Moreover, the first polynomial $p_1^\mathrm{iid}$ is of the same functional form as $p_1$, but $\alpha_\mathrm{iid}\neq\alpha$. 
In the higher orders, also the functional form of the polynomials $p_j^\mathrm{iid}$ is different from $p_j$; for example, $p_2^\mathrm{iid}$ contains the Hermite polynomials $H_4$ and $H_6$ while $p_2$ has an additional contribution from $H_2$. This can be understood as follows: an Edgeworth expansion is an expansion in terms of the cumulants $\kappa_{\ell}$ of the distribution. In the i.i.d.\ situation, the cumulants satisfy the scaling relation $\kappa_{\ell}[\cBN^\mathrm{iid}]=N^{1-\frac{\l}{2}}\kappa_{\ell}[\tilde{B}]$ for $\tilde{B}=B-\lr{\varphi,B\varphi}]$. In contrast, in the interacting case, each cumulant has a full series expansion, which leads to the additional contributions (see \cite{CLT} for a detailed discussion).

\subsection{Binding energy}

Another application of Theorem~\ref{thm:eigenstates:spectrum} concerns the binding energy, i.e., the energy it takes to remove one particle from the Bose gas in its ground state. Let us introduce the unscaled Hamiltonian
\begin{equation}
H(N,v) = \sum\limits_{j=1}^N\left(-\Delta_j+\Vext(x_j)\right)+ \sum\limits_{1\leq i<j\leq N}v(x_i-x_j).
\end{equation}
We now consider this Hamiltonian for $N$ particles and for $N-1$ particles, both with the same weak interaction $(N-1)^{-1} v =: \lambda_N v$, i.e., we consider the $N$-body Hamiltonian
\begin{equation}
H(N,\lambda_N v) = \sum\limits_{j=1}^N\left(-\Delta_j+\Vext(x_j)\right)+ \lambda_N \sum\limits_{1\leq i<j\leq N}v(x_i-x_j),
\end{equation}
which is the Hamiltonian from \eqref{HNtrap}, and the $(N-1)$-body Hamiltonian
\begin{equation}
H(N-1,\lambda_N v) = \sum\limits_{j=1}^{N-1} \left(-\Delta_j+\Vext(x_j)\right)+ \lambda_N \sum\limits_{1\leq i<j\leq N-1}v(x_i-x_j).
\end{equation}
If we denote the corresponding ground state energies by $E(N)$ and $\tilde{E}(N-1)$, the binding energy is defined as
\begin{equation}
\Delta E(N) := E(N) - \tilde{E}(N-1).
\end{equation}
Theorem~\ref{thm:eigenstates:spectrum} gives us an expansion of $E(N)$. But note that in our expansion we have not separated the contributions in $N$ coming from the number of particles and those from the coupling constant $\lambda_N=(N-1)^{-1}$. Hence, in order to obtain an expansion of $\tilde{E}(N-1)$, we need to replace in $E_N$ first the $N$ by $N-1$ and then $v$ by $\frac{N-2}{N-1}v$. The resulting series for $\tilde{E}(N-1)$ then needs to be rewritten as a power series in $N^{-1}$, just as in Theorem~\ref{thm:eigenstates:spectrum}. The result is a power series expansion of $\Delta E(N)$ in powers of $N^{-1}$.

\begin{theorem}\label{theorem_binding}
Under the assumptions of Theorem~\ref{thm:eigenstates:spectrum}, the binding energy $\Delta E(N)$ can be expanded as
\begin{equation}\label{main_expansion}
\Delta E(N) = \sum_{\ell=0}^a N^{-\ell} E^{\mathrm{binding}}_\ell + \mathcal{O}(N^{-(a+1)})
\end{equation}
for any $a \in \N$.
The coefficients $E^{\mathrm{binding}}_\ell$ are stated explicitly in \cite{binding_energy_reference}.
\end{theorem}

We know from \cite{grech2013} (or from Theorem~\ref{thm:eigenstates:spectrum} for $a=0$) that the leading order contribution is given by
\begin{equation}
E^{\mathrm{binding}}_0 = N \eH(v) - (N-1) \eH\big((N-2)(N-1)^{-1}v\big) = \eH + \frac{1}{2}\scp{\varphi}{\big( v*|\varphi|^2 \big)\varphi},
\end{equation}
where $\eH(v)$ is the Hartree energy with potential $v$. The next order $E^{\mathrm{binding}}_1$ was derived in \cite{nam2018} for the Bose gas on the torus. Note that \cite{nam2018} discusses the extension to the inhomogeneous case as a conjecture, which we address here with Theorem~\ref{theorem_binding} for $a=1$. For $a=2$ we compute the coefficient $E^{\mathrm{binding}}_2$ explicitly on the torus in \cite{binding_energy_reference}.

\section{Dynamics}\label{sec:dynamics}

\subsection{Two-body interaction}\label{sec:dynamics_two_body}
Let us assume that the Bose gas has initially been prepared in the ground state $\PsiN$ of $\HN$. Now we switch off the trap and let the gas propagate. Hence, the $N$-body wave function $\PsiN(t)$ at time $t>0$ is given by the solution of the time-dependent Schrödinger equation, generated by $\HN$ with $\Vext\equiv 0$. It is well known (see, e.g., \cite{hepp,ginibre1979,ginibre1979_2,SN1981,erdos2001,
bardos2000,frohlich2007_2,frohlich2009,rodnianski2009,knowles2010,pickl2011,chen2011,ammari2016,ammari2009,AN2011}) that the property of BEC is preserved by the time evolution, and that the time evolved condensate wave function $\pt$ is a solution of the Hartree equation,
\begin{equation}\label{hpt}
\i \partial_t \pt = \left(-\Delta + v*|\pt|^2 - \mpt\right)\pt\,,
\end{equation}
for some conveniently chosen phase $\mpt\in\R$.
The main result of \cite{QF} is an asymptotic expansion of the resulting dynamics.

\begin{theorem}\label{thm:dynamics}
Let $a\in\N_0$ and $t\in\R$. 
Then there exists $C(a)>0$ such that
\begin{equation}\label{eqn:thm:dynamics}
\Big\|\PsiN(t)-\sum\limits_{\ell=0}^a N^{-\frac{\ell}{2}}\psi_{N,\ell}(t)\Big\|_{L^2((\R^d)^N)}\leq \e^{C(a)t} N^{-\frac{a+1}{2}}\,.
\end{equation}
The coefficients $\psi_{N,\ell}(t)\in L^2_\sym((\R^d)^N)$ are given in \cite{QF} in full generality.
\end{theorem}
The leading order ($a=0$) of \eqref{eqn:thm:dynamics} was proven in \cite{lewin2015,mitrouskas2016}. Related results for the higher orders ($a>0$) were obtained in \cite{ginibre1979,ginibre1979_2,paul2019,corr}.
Theorem \ref{thm:dynamics} extends to a more general class of initial data. Besides, it implies an expansion of the reduced densities as well as a generalized Wick rule for the correlation functions  (see \cite{QF} for the full statement).

Analogously to \eqref{eqn:decomposition:PsiN}, the $N$-body wave functions $\Psi_{N,\ell}(t)$ are constructed by combining the time-evolved condensate $\pt$ with orthogonal excitations $\Chi(t)\in\Fock_{\perp\pt}$ and deriving a series expansion
\begin{equation}
\Big\|\Chi(t)-\sum_{\ell=0}^a N^{-\frac{\ell}{2}}\Chi_\ell(t)\Big\|_{\Fock_{\perp\pt}}\leq \e^{C(a)t}N^{-\frac{a+1}{2}}
\end{equation}
for the time-evolved excitations.
The leading order $\Chi_0(t)$ is given by the solution of the Bogoliubov equation, i.e., the time-dependent Schrödinger equation generated by the time-dependent analogue $\FockHz(t)$ of the leading operator in \eqref{intro:taylor}.
This is a  very useful approximation because the time evolution $U_0(t,t_0):\Fock_{\perp \varphi(t_0)} \to \Fock_{\perp \varphi(t)}$ generated by $\FockHz(t)$ acts as a Bogoliubov transformation. As a consequence, solving the Bogoliubov equation essentially reduces to the problem of solving a $2\times2$ matrix differential equation, which is a huge simplification in complexity compared to the full $N$-body problem.
Given the solution of the Bogoliubov equation, the first order correction is 
\begin{equation}\begin{split}\label{eqn:Chil:explicit:final}
\Chi_1(t)=&\sum_{j\in\{-1,1\}} \int_{\mathbb{R}^d} \dx  \,\mathfrak{C}_1^{(j)}(t;x)a^{\sharp_j}_x\Chiz(t)\\
&+\sum\limits_{(j_1,j_2,j_3)\in\{-1,1\}^{3}}
\int_{\mathbb{R}^{3d}} \dx^{(3)} \,
\mathfrak{C}^{(j_1,j_2,j_2)}_{3}(t;x^{(3)})a^{\sharp_{j_1}}_{x_1}a^{\sharp_{j_2}}_{x_2}a_{x_3}^{\sharp_{j_3}}
\Chiz(t)\,,
\end{split}\end{equation}
where we denoted $a^{\sharp_1}:=\ad$ and $a^{\sharp_{-1}}:=a$. The $N$-independent functions $\mathfrak{C}^{(j)}_1(t)\in L^2(\R^d)$ and $\mathfrak{C}^{(j_1,j_2,j_3)}_3(t)\in L^2((\R^{3})^d)$ are explicitly given in terms of the initial data and the solution of the $2\times 2$ matrix differential equation mentioned above (see, e.g., \cite[Equations~(3.22)]{proceedings}).

\subsection{Regularized Nelson model}

The techniques from the previous subsection can also be applied to non-relativistic quantum field models such as the regularized Nelson model in a many-particle mean-field limit. This model describes $N$ bosons that are linearly coupled to a quantized scalar (Klein--Gordon) field. The wave function $\Psi_N(t) \in L^2_{\rm{sym}}((\R^3)^N) \otimes \mathcal{F}$ evolves according to the Schr\"odinger equation with Hamiltonian
\begin{align}\label{eq:Nelson:Hamiltonian}
H_N^{\rm{Nelson}} =  \sum_{j=1}^N  \left(  - \Delta_j + N^{-1/2} \hat{\Phi}(x_j) \right) + \int_{\mathbb{R}^3} \d k \, \omega(k) a^{*}_k a_k .
\end{align}
Here, $\omega(k) = \sqrt{k^2 + m^2}$, with mass $m \geq 0$, is the dispersion relation of the field bosons, $a_k^{*}/a_k$ are the bosonic pointwise creation/annihilation operators, and 
\begin{align}
\hat{\Phi}(x) &= \int_{\mathbb{R}^3} \d k \, \frac{g(k)}{\sqrt{2\omega(k)}} e^{- 2 \pi i k x} \big( a_k^{*} + a_{-k} \big)
\end{align}
denotes the field operator with even cutoff function $g: \mathbb{R}^3 \rightarrow \mathbb{R}$ such that $g/\sqrt{\omega}$ and $g/ \omega$ are square integrable. If the particle-field state is initially prepared as a Bose--Einstein condensate of $N$ particles with condensate wave function $\varphi_0 \in L^2(\mathbb{R}^3)$ and a coherent state of field bosons with classical field $\sqrt N \alpha_0\in L^2(\mathbb R^3)$, then the condensation/coherent state structure is preserved under the time evolution generated by \eqref{eq:Nelson:Hamiltonian}, see \cite{F2012, AF2014,LP2018}. The corresponding mean-field equations describe the coupled evolution of the condensate wave function $\varphi(t)$ and  the classical field $\alpha(t)$. They are known as Schr\"odinger--Klein--Gordon equations and given by
\begin{align}
\begin{cases}
i \partial_t \varphi(t) &= \Big( - \Delta + \Phi(t) - \tfrac{1}{2} \langle\varphi(t) ,\Phi(t) \varphi(t) \rangle \Big) \varphi(t) 
\\
i \partial_t \alpha(t) &= \omega \alpha(t) +  \tfrac{g}{\sqrt{2\omega}} \widehat{|\varphi(t)|^2}
\\
\Phi(t,x) &= \int_{\mathbb{R}^3} dk \, \tfrac{g(k)}{\sqrt{2\omega}(k)} e^{- 2 \pi i k x} \left( \overline{\alpha(t,k)} + \alpha(t,-k) \right) 
\end{cases}
\end{align}
with initial datum $(\varphi_0,\alpha_0)$, where $\widehat{|\varphi(t)|^2}$ is the Fourier transform of $|\varphi(t)|^2$.
In \cite{FLMP2021} it was shown that (for suitably chosen initial states) the time evolution of the regularized Nelson model satisfies an asymptotic expansion in the spirit of Theorem~\ref{thm:dynamics}. The main difference to the previous subsection, where $N$ bosons interact via pair potentials, is that the system consists now of two types of particles. In order to study the fluctuations around the mean-field dynamics it is therefore necessary to factor out the Bose--Einstein condensate as well as the coherent state of the field bosons, which can be done in a similar manner as in \eqref{eqn:decomposition:PsiN}. The resulting orthogonal excitations $\Chi(t)$ are elements of the double Fock space $\Fock_{\perp\pt} \otimes \mathcal{F}$ and the corresponding quadratic Bogoliubov Hamiltonian $\mathbb H_0(t)$ (and its higher-order corrections) are operators on this space.
The Bogoliubov dynamics captures not only correlations among the particles and  the field excitations themselves but also between the particles and field excitations.
\subsection*{Acknowledgements}
It is our pleasure to thank Marco Falconi, Nata\v{s}a Pavlovi\'c, Peter Pickl, Robert Seiringer and Avy Soffer for the collaboration on the works \cite{corr,QF,spectrum,FLMP2021,LP2018,mitrouskas2016}. L.B.\ was supported by the  German Research Foundation within the Munich Center of Quantum Science and Technology (EXC 2111).
N.L.\ acknowledges support from the Swiss National Science Foundation through the NCCR SwissMap and funding from the European Union's Horizon 2020 research and innovation programme under the Marie Sk\l odowska-Curie grant agreement N\textsuperscript{o} 101024712.

\fontsize{10}{12}\selectfont

\renewcommand\refname{} 
\begingroup
\let\section\subsubsection
\makeatletter
\renewcommand\@openbib@code{\itemsep\z@}
\makeatother
\bibliographystyle{abbrv}
\bibliography{bib_file}
\endgroup

\normalsize

\end{document}